\documentclass[12pt]{article}

\usepackage{fancyhdr}
\usepackage{amsmath}
\usepackage{amsmath}
\usepackage{latexsym}
\usepackage{amssymb}
\usepackage{amsfonts}
\usepackage[english]{babel}
\usepackage{biblatex}
\usepackage[dvips]{graphicx}
\usepackage{fullpage}
\usepackage{graphicx}
\usepackage{mathrsfs}
\usepackage{tabularx}
\usepackage{caption}
\usepackage{subcaption}

\newcommand{\abs}[1]{\lvert#1\rvert}

\title{Proposal of the Electrically Charged Stellar Black Holes as Accelerators of Ultra High Energy Cosmic Rays}
\author{Jose Soto-Manriquez \\
	    Instituto de F\'isica, Benem\'erita Universidad Aut\'onoma de Puebla \\
		Apartado Postal J-48, Puebla, Pue., 72570, M\'exico \\
	    \texttt{jsoto@ifuap.buap.mx}}
			\date{ }
\begin{document}

\maketitle

\begin{abstract}
A new mechanism for the acceleration of ultra high energy cosmic rays (UHECRs) is presented here. It is based on the tunnel-ionization of neutral atoms approaching electrically charged stellar black holes and on the repulsion of the resulting positively charged atomic part by huge, long range electric fields. Energies above $10^{18}$ eV for these particles are calculated in a simple way by means of this single-shot, all-electrical model. When this acceleration mechanism is combined with the supernova (SN) explosions in the galactic halo of the massive runaway stars expelled from the galactic disk, then the various results obtained are shown to be compatible with virtually all the main observational facts about these UHECRs. Among these facts, this model predicts nearly the correct values of the measured top energy of the UHECRs and their flux in a specified EeV energy range. It also explains the near isotropy of arrivals of these energetic particles to Earth, as it has recently been measured by the Auger Observatory.
\end{abstract}

\section{Introduction}
\label{sec.1}

Cosmic rays are very energetic particles traveling in all directions in outer space. Their energies vary from a few $10^{7}$ eV to beyond $10^{20} eV$. The most extensively studied and cited model for the creation of intragalactic cosmic rays is known as the \textit{diffuse shock acceleration} (DSA) in the expanding outer shock fronts of SN remnants. This model was originally proposed by Enrico Fermi in 1949 and has been improved since then. This mechanism can presumably raise a small fraction of nuclei up to around $10^{15}$ eV. However, many astrophysicists think that it is unlikely that a SN could produce cosmic rays up to $10^{18}$ eV through this process.

The identification of the sources of UHECRs with energies above $10^{18}$ eV has remained one of the greatest challenges in Astrophysics for at least five decades. The multitude of theories conceived with the purpose of explaining the origin of the UHECR are broadly categorized into the ``bottom-up" (BU) and the ``top-down" (TD) scenarios.

In the BU scenario, charged particles become UHECRs through powerful acceleration processes having a conventional nature. The best compact candidates are quasars, Gamma Ray Bursts (GRBs), Active Galactic Nuclei (AGN), rotating neutron stars, etc., while the best non-compact candidates are radio loud galaxies and colliding galaxies.

In the TD scenario, the charged particles are not accelerated as such. The UHECR is directly produced via a decay cascade of some supermassive hypothetical particles which are not even describable by the Standard Model of the elementary particles. However, \textbf{both scenarios are incompatible with the observational facts} collected in the last decade; in particular, with those of the Auger Observatory, and the situation remains largely unsettled \cite{1, 2}.

The mechanism that I propose here for the acceleration of UHECR \textbf{rests on the existence of positively charged, mostly quiescent, stellar black holes}. Since it has been theoretically established that black holes can be characterized by three parameters: its mass $M_{bh}$, its electric charge $Q_{bh}$ and its angular momentum $J_{bh}$, the two assumptions made in this model are \textit{firstly} that indeed these astrophysical objects are capable of acquiring enormous initial charges and \textit{secondly} that this charge is not immediately neutralized by the environment, as is generally assumed, and that it can survive for a large number of years. This last assumption is made in order to explain the relatively stable average flux of UHECRs in space, in spite of the low birthrate of black holes, as determined by the average rate of a few SN explosions per century in our galaxy and in most of the neighboring galaxies. We will see that on the basis of these two assumptions, the new acceleration scheme leads to predictions which are compatible with several important observational facts and, in a couple of key cases, it is quantitatively close to them.

Qualitatively, the process is as follows: a neighboring neutral atom is gravitationally attracted by the charged black holes (CBH) subjecting it to an increasingly powerful electric field, which tends to ionize the atom through the quantum tunneling effect. Once the atom is singly ionized through this process, the positive part suddenly acquires a huge repulsive potential energy, with enormous radial accelerating electric fields lasting for hundreds of thousands of kilometers, while the electron is powerfully attracted by the CBH, who suffers an elementary discharge. Thus, according to this model, the acceleration process takes place on a one-by-one basis, without the involvement of any cosmic violence, galactic dimensions or of time-changing magnetic fields of any nature.

\section{Theory}
\label{sec.2}

\subsection{Statement}

One of the most general stationary, black hole solutions for the general relativity equations, which is known as the \textit{Kerr-Newman metric}, dates from 1965. In this solution, the black hole is described as possessing both charge and angular momentum. While the mass of the black hole can take any positive value, the charge $Q_{bh}$ and the angular momentum $J_{bh}$, when expressed in Planck units, are constrained to satisfy the following expression \cite{3}:
\begin{equation}
Q_{bh}^{2} + \left( \frac{J_{bh}}{M_{bh}} \right)^{2} \leq M_{bh}^{2}
\label{eq.1}
\end{equation}
The black holes that saturate this inequality are known as \textit{extremal}. The solutions of Einstein's equations violating this inequality are deemed unphysical. The upper bound $Q_{bh}^{Max} = M_{bh}$ of the allowed positive charge is found for the extremal solution when the CBH does not rotate. I interpret this last equation as indicating that the number of Planck units of charge $q_{pk}$ in $Q_{bh}^{Max}$ is equal to the number of Planck units of mass $m_{pk}$ in $M_{bh}$; that is,
\begin{displaymath}
\frac{Q_{bh}^{Max}}{q_{pk}} = \frac{M_{bh}}{m_{pk}}
\end{displaymath}
When $Q_{bh}^{Max}$ is expressed in C and $M_{bh}$ in kg, and with the equivalences $q_{pk} = 1.8755 \times 10^{-18}$ C and $m_{pk} = 2.1764 \times 10^{-8}$ kg, I obtain the following expression:
\begin{displaymath}
Q_{bh}^{Max} = 8.6174 \times 10^{-11} M_{bh}  \mbox{ C-kg}^{-1}.
\end{displaymath}
As an example, for $M_{bh} = 10 \mbox{M}_{\odot}$, with $\mbox{M}_{\odot} = 1.9707 \times 10^{30}$ kg, the electric charge becomes $Q_{bh}^{Max} = 1.7 \times 10^{21}$ C.

Although this new acceleration model is applicable to the different types of neutral atoms, in this part of the work, I deal in detail with the ionization of the Hydrogen atom, since for this element, all the required formulae are readily available. 

In the first place, I consider an H atom that at a radial distance $r \rightarrow \infty$ moves towards the CBH with a speed $v_{orig}$, possessing the energy $U = 1 / 2 m_{H} v_{orig}^{2}$. At a still large, but much closer distance $r_0$, and with an impact parameter $S \ll r_{0}$, the gravitational pull increases the speed to 
\begin{displaymath}
v_{0} \approx v_{orig} + \left( \frac{2G M_{bh}}{r_{0}} \right)^{1/2}
\end{displaymath}
with the velocity vector pointing closely towards the force center. The associated angular momentum is given by
\begin{displaymath}
L = m_{H} v_{0} S
\end{displaymath}
Due to the conservation of the angular momentum, the initial impact parameter is given by
\begin{displaymath}
S_{orig} = \frac{S v_{0}}{v_{orig}}
\end{displaymath}
In these expressions, $m_{H}$ represents the mass of the Hydrogen atom, while $G$ represents the Newtonian constant of gravitation.

For $U > 0$, the atomic trajectory will be of a hyperbolic type. However, for the present problem, it is not necessary to obtain the solution for the whole trajectory. It suffices to obtain it only for the radial distance $r(t)$. This can be accomplished by solving the differential equation \cite{4}:
\begin{equation}
m_{H} \ddot{r} - \frac{L^{2}}{m_{H} r^{3}} = \frac{G m_{H} M_{bh}}{r^{2}}
\label{eq.2}
\end{equation}
which satisfies the initial conditions $r(0) = r_{0}$ and $\dot{r} (0) \approx - v_{0}$.

Once the solution for $r(t)$ is found, by simple substitution it is possible to obtain the corresponding functions for the electric strength $\mathcal{E} (t)$ acting on the atom:
\begin{equation}
\mathcal{E} (t) = \frac{Q_{bh}}{4 \pi \epsilon_{0} r^{2} (t)}
\label{eq.3}
\end{equation}
For the rate of tunnel ionization of a Hydrogen atom exposed to an external electric field $\mathcal{E} (t)$, I use Landau's formula \cite{5}. In SI units, this formula is given by:
\begin{equation}
\mathcal{W}_{L} (t) = 4 \omega_{a} \left( \frac{\mathcal{E_{B}}}{\abs{\mathcal{E} (t)}} \right) \mbox{ Exp } \left( \frac{- 2 \mathcal{E}_{B}}{3 \abs{\mathcal{E} (t)}} \right)
\label{eq.4}
\end{equation}
where $\omega_{a} \approx 4.1341 \times 10^{16} \mbox{ s}^{-1}$ corresponds to twice the ionization angular frequency of the Hydrogen atom and $\mathcal{E}_{B} = 5.1422 \times 10^{11}$ V/m is the magnitude of the electric field produced by a proton at the distance of the Bohr radius $a_{0}$. The Landau rate formula is considered to be accurate when $\abs{\mathcal{E} (t)} \ll \mathcal{E}_{B}$ \cite{6}.

If the probability of the atom for staying neutral along its trajectory is represented by $\mathcal{P} (t)$, then its rate of change will be proportional to its current value, with a proportionality factor given by the instantaneous Landau ionization rate, in the diminishing sense:
\begin{equation}
\frac{d \mathcal{P} (t)}{dt} = - \mathcal{W}_{L} (t) \mathcal{P} (t)
\label{eq.5}
\end{equation}
This equation is similar in structure to the one that describes the occupation probability of the excited atomic state in spontaneous radiative decay, although the mechanisms for the probability depletion rates of the atomic states are of different natures.

With $\mathcal{W}_{L} (t)$ being already a known function of time and with the initial condition at $r_{0}$ given by $\mathcal{P} (0) = 1$, the integration of equation \ref{eq.5} gives the function:
\begin{equation}
\mathcal{P} (t) = \mbox{ Exp } \left( - \int_{0}^{t} \mathcal{W}_{L} (\tau) d \tau \right)
\label{eq.6}
\end{equation}
Finally, since the electron is either bound or ionized, the actual ionization rate $\mathcal{I} (t)$ must be equal to $- d \mathcal{P} (t) / dt$, and the following function is obtained:
\begin{equation}
\mathcal{I} (t) = \mathcal{W}_{L} (t) \mathcal{P} (t)
\label{eq.7}
\end{equation}
Upon the ionization of the H atom at the radial distance $r(t)$, the proton instantaneously acquires a huge repulsive potential energy given in eV by
\begin{equation}
W_{eV} (t) = 6.2415 \times 10^{18} \frac{Q_{bh} q_{p}}{4 \pi \epsilon_{0} r(t)} \mbox{ eV/J}
\label{eq.8}
\end{equation}
In this work, the differential equation \ref{eq.2} was solved numerically by means of the program $\mbox{MATHEMATICA}^{\circledR}$. This solution consists of a highly accurate interpolation function for $r(t)$, valid within a specified temporal range of integration. This function is differentiable a number of times and is also integrable. As a consequence, $\mathcal{E} (t)$, $\mathcal{W}_{L} (t)$, $\mathcal{P} (t)$, $\mathcal{I} (t)$, and $W_{eV} (t)$ are also given by their corresponding interpolation functions.

\subsection{Applications of the Theory}

In this section, I present the most illustrative graphical results of the application of the model to one specific case of CBH. This case corresponds to a stellar object with mass $M_{bh} = 10 \mbox{ M}_{\odot}$ and electric charge $Q_{bh} = \, + 10^{18}$ C. This charge is close to $Q_{bh}^{Max} / 1700$. I have also found that it is convenient to express the radial distance in terms of
\begin{displaymath}
r_{B} = a_{0} \sqrt{\frac{Q_{bh}}{q_{p}}}.
\end{displaymath}
This is the distance at which the charge $Q_{bh}$ produces an electric field equal to $\mathcal{E}_{B}$. For $Q_{bh} = 10^{18}$ C, I obtain $r_{B} = 1.322 \times 10^{5}$ km.

Equation \ref{eq.2} was solved for a number of impact parameters which are expressed as integer multiples of $r_{B}$:
\begin{equation}
S_{j} = (j - 1) r_{B}, \hspace{5cm} \mbox{ for } j = 1, 2, ..., N_{ip}
\label{eq.9}
\end{equation}
with $N_{ip}$ being sufficiently large until no significant ionization probability occurs. For the present case, it has been found that $N_{ip} = 100$ is an adequate value. The solutions obtained are represented as $r (t, j)$. Consequently, the other functions are represented by the symbols $\mathcal{E} (t, j)$, $W_{eV} (t, j)$, $\mathcal{W}_{L} (t, j)$, $\mathcal{P} (t, j)$, and $\mathcal{I} (t, j)$, respectively.

\begin{figure}[!h]
  \centering
  \includegraphics[width=6.5in]{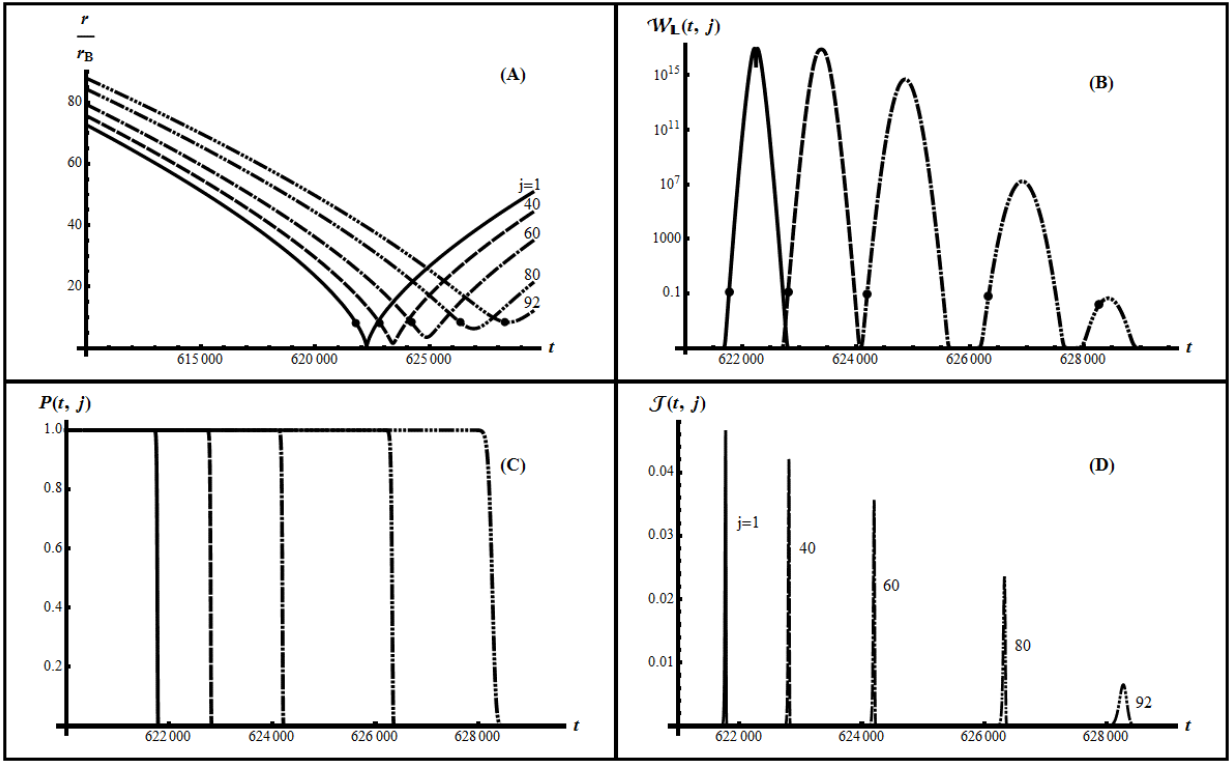}
\caption{Plots of the functions $r (t, j)$, $\mathcal{W}_{L} (t, j)$, $\mathcal{P} (t, j)$, and $\mathcal{I} (t, j)$ for $j = 1$, 40, 60, 80, and 92, in the neighborhood of the CBH. The dots in the curves of parts (A) and (B) give the values of these functions when the ionizations take place in the peaks of part (D).}
\label{fig.1}
\end{figure}

\begin{figure}[!h]
  \centering
  \includegraphics[width=6in]{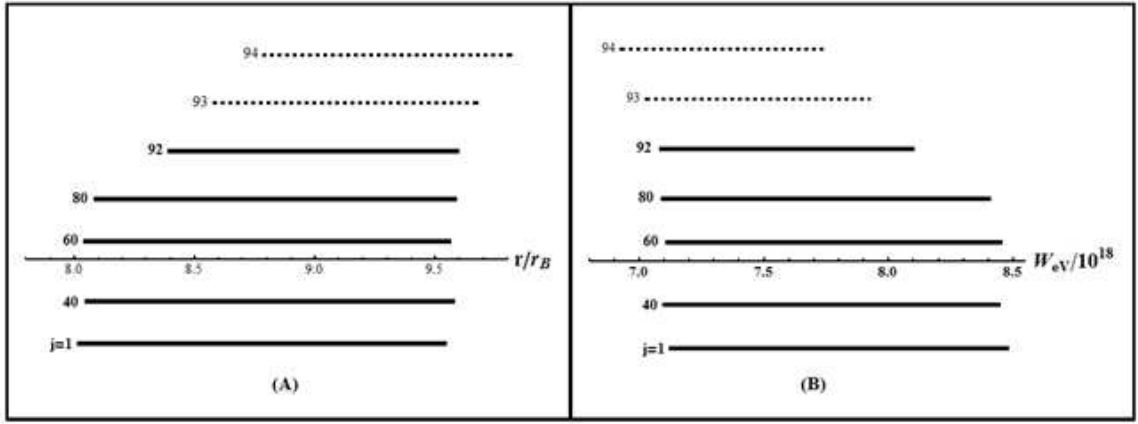}
  \caption{(A) represents the radial ranges, while (B) the proton energy ranges where the ionization of the H atom can take place with certainty ($j = 1$, 40, 60, 80, and 92) or with partial probability ($j = 93$ and 94), as given by the limits of the minimum integration range of $\mathcal{I} (t, j)$.}
\label{fig.2}
\end{figure}

Of the $N_{ip}$ obtained, only those functions corresponding to $j = 1, 40, 60, 80$, and 92 are shown in figures \ref{fig.1} and \ref{fig.2}. Although in figure \ref{fig.1}, the composed plots of $r (t, j)$, $\mathcal{W}_{L} (t, j)$, $\mathcal{P} (t, j)$, and $\mathcal{I} (t, j)$ appear in the order that they are obtained, I will describe them starting with the last group.

According to the calculations, the ionization rate functions $\mathcal{I} (t, j)$ appear as sharp temporal peaks. Starting at $r_{0} = 10^{3} r_{B}$ and with $v_{orig} = 1000$ m/s, it takes the H atoms around one week to reach these peak rates with timing spreading in a range close to 1.86 hours. However, the full duration of the $j = 1$ peak is close to 133 seconds, while that of the $j = 92$ peak is close to 570 seconds.

The dots appearing in the plots for $r (t, j)$ and $\mathcal{W} (t, j)$ in figure \ref{fig.1} are the values of these functions at the times when the peaks of their respective functions $\mathcal{I} (t, j)$ occur. As can be seen in figure \ref{fig.1} (A), all the ionizations take place at a radial distance close to 8.8 $r_{B}$ After its ionization point, each curve $r (t, j)$ is meaningless for the H atom and must be replaced by an almost vertical straight line, starting at the ionization point, to represent the fast proton being repelled by the CBH.

On the other hand, from the logarithmic plots of figure \ref{fig.1} (B), we can see that the ionization peaks occur approximately in those regions where $0.013 \mbox{ s}^{-1} \lesssim \mathcal{W}_{L} (t, j) \lesssim 0.3 \mbox{ s}^{-1}$, in spite of the fact that for values of $j$ lower than 40, the functions $\mathcal{W}_{L} (t, j)$ can reach peak values as high as $8 \times 10^{16} \mbox{ s}^{-1}$, although for $j = 92$, the peak value is barely $0.04 \mbox{ s}^{-1}$. The probability of ionization, as given by the integral of $\mathcal{I} (t, j)$ is equal, or very close to 1 up to $j = 92$. For higher values of $j$, the probability of ionization of the H atom rapidly decays to a fraction of one and then to negligible values. Thus, the maximum value of $j$ for which complete ionization occurs is $j_{max}^{(H)} \approx 92$ and at the initial distance $r_{0}$, the \textit{cross section} of this CBH for the ionization of H atoms is given by
\begin{displaymath}
\sigma_{H} \approx \pi \left[ \left( j_{max}^{(H)} - 1 \right) r_{B} \right]^{2}.
\end{displaymath}

From figure \ref{fig.1} (C), we can see that fast transitions from 1 to 0 of the survival probabilities $\mathcal{P} (t, j)$ of the H atoms are coincident in time with the appearance of the peaks of their respective functions $\mathcal{I} (t, j)$.

The ionization rate peaks of $\mathcal{I} (t, j)$ also span some radial distance intervals and some energy ranges of the released protons. The procedure that I followed in most cases to obtain them consists in finding, by trial an error, the accurate values of the maximum lower limit $t_{min(j)}$ and the minimum upper limit $t_{max(j)}$ of the time integral of $\mathcal{I} (t, j)$, which still gives a value equal to 1. These limits are substituted in $r (t, j)$ and in $W_{eV} (t, j)$ for finding the extremes of their respective intervals. The case for $j = 92$ is slightly different because $t_{max(92)}$ occurs in the receding part of the atomic trajectory and the functions $r(t, 92)$ and $W_{eV} (t, 92)$ must be evaluated at the lower integration limit $t_{min(92)}$ and at the time $t_{close(92)}$ of the closest approach of this trajectory to the CBH in order to obtain the correct radial and energy ranges of the proton.

The results of these calculations are shown in figure \ref{fig.2}. Each of the horizontal black segments in part (A) indicates the radial extent where the ionization of the H atom can take place with certainty, for a given value of $j$. This segment is given in terms of $r_{B}$, and no ionization caused by the tunneling mechanism can take place outside of it. From this figure, we can deduce that virtually all the possible ionizations take place inside a spherical shell centered at the CBH, of average radius $r_{sh} \approx 8.8 r_{B}$ and thickness $\Delta r_{sh} \approx 1.6 r_{B}$. This spherical shell is reached by the Hydrogen atom almost frontally for small values of $j$, obliquely for intermediate values, and almost tangentially for the largest ionizing values of $j$.

Even more interesting is the information provided by figure \ref{fig.2} (B). In this figure, each of the black segments indicates the energies in eV available to the protons, once the H atom is tunnel-ionized, for the given values of $j$. Here, we can see that they are contained within a range of $1.4 \times 10^{18}$ eV, approximately, and are centered near $7.8 \times 10^{18}$ eV. This is perhaps the most important piece of information provided by this model for the chosen CBH and it is fully compatible with the values of energy measured for UHECRs.

Once a proton is freed from its electron, it is immediately subjected to the electric field $\mathcal{E} [r_{sh}] \approx 6.64 \times 10^{9}$ V/m of the CBH, and between $r_{sh}$ and $2 r_{sh}$, which is about $1.163 \times 10^{6}$ km, the average of the accelerating field is $< \mathcal{E} > = \mathcal{E} [r_{sh}] / 2$. This clearly represents a viable embodiment of the much sought after \textit{``cosmic accelerator"}.

As was already mentioned, all the above calculations were done with $v_{orig} = 1,000$ m/s. I have also repeated all these calculations with $V_{orig} = 500$, 5,000 and 10,000 m/s, obtaining the same curves for the functions $\mathcal{E} (t, j)$, $W_{eV} (t, j)$, $\mathcal{W}_{L} (t, j)$, $\mathcal{P} (t, j)$, and $\mathcal{I} (t, j)$, but at displaced times: Increased for the first case, decreased for the second case and further decreased for the third case. However, no discernible difference is observed for the two types of plots of figure \ref{fig.2} among all the values of $v_{orig}$ considered.

\subsection{Other values of $Q_{bh}$}

A systematic study with other values of charge for the CBH was done in detail, but only the important numerical results are presented in table \ref{tab.1}. The values of $Q_{bh}$ range from $10^{15}$ to $10^{21}$ Coulombs. Other combinations of $M_{bh}$ for each of these values of $Q_{bh}$ were also tried, but the results show only minor differences with those obtained with its current value of 10 $\mbox{M}_{\odot}$.

\begin{table} [h]
\begin{center}
\begin{tabular}{|c|c|c|c|c|c|c|}
\hline
$\mathbf{Q_{bh} (C)}$   & $\mathbf{r_{B} (10^{3} \mbox{ km})}$   & $\mathbf{r_{sh} / r_{B}}$ & $\mathbf{\Delta r_{sh} / r_{B}}$  & $\mathbf{W_{eV} (\mbox{EeV})}$  & $\mathbf{\Delta W_{eV} (\mbox{EeV})}$ \\
\hline
$10^{15}$   & 4.1807    & 8.349 & 1.651 & 0.260 & 0.051 \\
$10^{16}$   & 13.220    & 8.637 & 1.887 & 0.796 & 0.174 \\
$10^{17}$   & 41.807    & 8.755 & 1.749 & 2.480 & 0.495 \\
$10^{18}$   & 132.22    & 8.806 & 1.569 & 7.782 & 1.386 \\
$10^{19}$   & 418.07    & 9.177 & 1.808 & 23.655 & 4.659 \\
$10^{20}$   & 1322.0    & 9.288 & 1.837 & 73.918 & 14.618 \\
$10^{21}$   & 4180.7    & 9.364 & 1.684 & 231.446 & 41.636 \\
\hline
\end{tabular}
\caption{Main results for CBHs with $Q_{bh}$ ranging from $10^{15}$ to $10^{21}$ C with the same mass $M_{bh} = 10 \mbox{ M}_{\odot}$.}
\label{tab.1}
\end{center}
\end{table}

Apart from the distances $r_{B}$, which depend directly on $Q_{bh}$, the other quantities tabulated are the radius and thickness of the ionization spherical shell region, expressed in terms of the distance $r_{B}$, the central energy and the energy spread of the proton cosmic rays, expressed in terms of $10^{18}$ eV (EeV).

There are some regularities which can be observed in this table, mostly because the distance $r_{B}$ is proportional to $\sqrt{Q_{bh}}$, in all the cases, the radius of the ionization shell remains close to 8.8 $r_{B}$ and after changing the value of the charge by six orders of magnitude, this relation increases only by $\sim 12 \%$. Although the thickness of the shell does not show a regular behavior, its value also remains close packed around 1.74 $r_{B}$ after these big changes.

Another valuable information obtained from the table, which is directly related to the previous one, is the dependence of the central energy $W_{eV}$ of the accelerated proton with the charge $Q_{bh}$. In it, we can observe that in order to increase this energy by one order of magnitude, the charge must increase by two orders of magnitud. This may explain the observational fact that the UHECR's are increasingly more difficult to observe as their energy increases, since after a SN explosion, the very large values of the initial charge $Q_{bh}$ must be much less probable than small values. Thus, the charge for events of the order of $10^{20}$ eV must be about $10^{4}$ times larger than the charge required for the more frequent events of $\sim 10^{18}$ eV.

\section{More compatibilities with observations}
\label{sec.3}

\subsection{Other atoms}
\label{sec.3.1}

Attention is now given to UHECR's consisting of nuclei belonging to atoms different from Hydrogen. These atoms are represented here by the generic symbol $\xi$. It is assumed that for each H atom at the distance $r_{0}$ from the CBH there is a fraction $n_{\xi}$ of atoms of the $\xi$-type also moving towards the BH with the same velocity $v_{0}$ and with the same type of trajectories as those of the H atoms. As the $\xi$-atom becomes sufficiently close to the CBH, it is \textit{singly ionized} through the quantum tunneling effect. As soon as this atom becomes a positive ion, it is strongly repelled by the CBH and very rapidly reaches long distances from it, with basically a \textit{null probability} of undergoing a second tunnel ionization, and eventually acquiring an enormous kinetic energy. As the very energetic ion travels through space, it can lose the rest of its electrons through collisions with atoms in the interstellar medium, thus becoming a $\xi$-nucleus UHECR, like those that reach the Earth.

Since I am not aware of the existence of the tunneling rate functions $\mathcal{W}_{L}^{( \xi )} [ \mathcal{E} ]$, I cannot include here detailed calculations for other atoms as those done for the H atom. However, it is to be expected that, once they are known, each one of them will also exhibit ionization rate peaks in ranges of values of the type $\epsilon_{min}^{( \xi )} < W_{L}^{( \xi)} [ \mathcal{E} ] < \epsilon_{max}^{( \xi )}$, with $\epsilon_{min}^{(\xi)} < \epsilon_{max}^{(\xi)} < (\mbox{a few s}^{-1})$, as well as a spherical ionization shell of average radius $r_{sh}^{(\xi)}$ and thickness $\Delta r_{sh}^{(\xi)}$, with a corresponding peak potential energy given by $W_{eV} [r_{sh}^{(\xi)}]$. It is also to be expected that at the initial distance $r_{0}$ there will be a maximum impact parameter
\begin{displaymath}
S_{max}^{(\xi)} = \left ( j_{max}^{(\xi)} - 1 \right) r_{B},
\end{displaymath}
beyond which no complete ionization probability will occur. At this initial distance, the CBH will present a cross section for this $\xi$-atom given by
\begin{displaymath}
\sigma_{\xi} = \pi \left[ \left( j_{max}^{(\xi)} - 1 \right) r_{B} \right]^{2}.
\end{displaymath}
 By defining the \textit{abundance gain factor} as 
\begin{displaymath}
\mathscr{G}_{\xi} = \frac{\sigma_{\xi}}{\sigma_{H}},
\end{displaymath}
it follows that for every proton accelerated by this mechanism, there are approximately $\mathscr{G}_{\xi} n_{\xi}$ ions of the $\xi$ atoms accelerated by the same process. Thus, \textit{for $j_{max}^{(\xi)} > j_{max}^{(H)}$, as is expected for atoms with weakly bound outer electrons, there will be a net increase in the abundance for the $\xi$-CR relative to that of the proton-CR as compared to the abundance of the $\xi$-atoms relative to that of the H atoms in the interstellar environment of the CBH.}

This prediction is indirectly confirmed by the experimentally known fact that the abundances of the CRs accelerated \textit{by the shock fronts} of SN remnants are controlled by atomic parameters such as the \textit{first ionization potential} (FIP) and not by nucleosynthetic processes that take place in SN explosions: the lower FIP chemical elements being systematically favored relative to the higher FIP ones \cite{7}. Thus far, this phenomenon has remained without a scientific explanation.

Hence, from this new acceleration model, a methodology emerges for calculating the different abundance gain factors $\mathscr{G}_{\xi}$ (not their compositional abundances arriving at the Earth), once the corresponding tunneling rate functions $\mathcal{W}_{L}^{(\xi)} [\mathcal{E} ]$ are known. Obtaining gain factors numerically consistent with measurements would represent a major test for this new acceleration model.

It must be pointed out, however, that there is the possibility for the gain factors $\mathscr{G}_{\xi}$ of being nearly constants, almost independent of the values of $M_{bh}$ and $Q_{bh}$. Thus, for the values of $M_{bh} = 5$, 10 and 15 $\mbox{M}_{\odot}$ in all the allowed combinations with $Q_{bh} = 10^{n}$ C (for the integer $n$ running from 15 up to 21) I have computationally determined the values of the parameters $j_{max}^{(H)}$ that still make the ionization probabilities equal to one. In all cases, their values remained very close to 92. This being in spite of the six orders of magnitude change of $Q_{bh}$. Again, this is most probably due to the charge dependence of $r_{B}$.

Because the tunnel ionization processes of the H and of the $\xi$-atoms have a common nature, it is likely that the parameter $j_{max}^{(\xi)}$ for the $\xi$-atom will also show small variability with respect to changes in $M_{bh}$ and $Q_{bh}$. If this proves to be the case, then the value of the enrichment factor
\begin{displaymath}
\mathscr{G}_{\xi} = \left( \frac{j_{max}^{(\xi)} - 1}{j_{max}^{(H)} - 1} \right)^{2}
\end{displaymath}
will be nearly the same for all the valid combinations of these two quantities.

However, it is only through full complete simulations for the $\xi$-atoms, similar to those done for the H-atom, and by using highly accurate tunneling rate functions $W_{L}^{(\xi)} (\mathcal{E})$ that we can determine the exact behavior of $\mathscr{G}_{\xi}$ for different cases of CBHs.

\subsection{Top energy limit of UHECRs}
\label{sec.3.1.2}

According to this new model, the highest possible energy for these particles will be dictated by the mass of the most massive stellar black hole near us. So far, the \textit{most massive stellar black hole} that has been measured is in the eclipsing X-ray binary known as M33 X-7, in the nearby Triangle galaxy \cite{8}. The presence of its small companion has allowed the measurement of its mass as $M_{bh} \approx 15.7 \mbox{ M}_{\odot}$. However, I think it is reasonable to assume that this \textit{mass size} can also be reached by the largest \textit{isolated} stellar BHs in our galaxy, which are basically non-measureable.

By using the formula $Q_{bh}^{Max} = 8.6174 \times 10^{-11} M_{bh} \mbox{ C-kg}^{-1}$, the value of the maximum charge $Q_{bh}^{Max} = 2.67 \times 10^{21}$ C is obtained for this top mass. For this charge, $r_{B} = 6.83 \times 10^{9}$ m, an since in this case, $r_{sh} \approx 9.37 r_{B}$, then $r_{sh} \approx 6.4 \times 10^{10}$ m. When these values are substituted in the expression:
\begin{displaymath}
W_{eV}^{Max} = 6.2415 \times 10^{18} \frac{Q_{bh}^{Max} q_{p}}{4 \pi \epsilon_{0} r_{sh}} \frac{\mbox{eV}}{\mbox{J}},
\end{displaymath}
\textbf{the upper limit} $\mathbf{W_{eV}^{Max} = 3.75 \times 10^{20}}$ \textbf{is obtained for the UHECRs energy. This limit is just 17 \% higher than the highest energy of} $\mathbf{3.2 \times 10^{20}}$ \textbf{eV for a recorded event, which was measured in 1993 by the Fly's eye system.} This top energy is indicated by the dashed vertical arrow in the experimental cosmic ray spectrum showed in figure \ref{fig.3}.

This result \textit{seems to be closer to an experimental confirmation} of the new acceleration model than just a simple compatibility with measurements, and consequently points towards a galactic origin for the majority of the most energetic UHECRs reaching our planet.

\section{The onset of the acceleration process}
\label{sec.4}

Thus far, the assumptions of the \textit{existence of the BH charge} and of its \textit{long-term durability} have already led to a number of important compatibilities with observations. Although in this paper, they are technically to be taken as working hypotheses, it is still possible to advance some preliminary arguments in favor of the second one.

This is in spite of the fact that in the astrophysical literature, there is a general consensus that, even if due to some unknown mechanism the BHs were produced with large net charges, they would be rapidly discharged due to the ionized interstellar space that surrounds them; see reference \cite{9}, for example. However, it is my view that there exists the possibility for this rapid neutralization of not taking place.

The blast wave of the ejected material moves with a radial speed initially higher than 1/25 of the speed of light and as it expands, it picks up essentially all the atoms and molecules it encounters in the space surrounding the exploded star. Besides that, the magnetic shock front that accompanies this ionized blast wave accelerates as cosmic rays the electrons, protons, and ionized atoms that it encounters \cite{10}. Thus, inside this expanding wave, apart from the compact stellar object, a bubble filled by the flow of a very low density gas is left, which is also ejected by the star explosion, whose radial speed is nearly proportional to the radial distance to the center. Moreover, even if the initial explosion is not spherical, its blast wave becomes spherical at more advanced stages, after it has swept up several times its initial mass from the interstellar medium. \textbf{In this way, the CBH could be isolated from the interstellar medium, preventing in this way its rapid discharge.}

According to \cite{10}, the blast wave expansions can last for many thousand years, and even for more than $10^{5}$ years \cite{11}, until the blast front is diluted in the interstellar medium. Then, there exits the possibility that \textit{once the wave front has been sufficiently weakened, the external neutral atoms will be capable of leaking to its interior before the electrons can do it.} Subsequently, these atoms would be gravitationally attracted by the BH, marking in this way the onset of the process that leads to the emission of UHECRs through this acceleration mechanism proposed in this work.

However, it is more difficult to identify the mechanism by which the BHs could acquire such large \textit{Kerr-Newman's} electric charges. Nevertheless, some considerations about this problem are made in the supplement.

\section{Additional compatibilities}
\label{sec.5}

In order to arrive to other compatibilities, with the experimental observations, it is necessary to extend the acceleration model as to include the role of the \textit{massive runaway stars} in our galaxy. Among these additional observations are the \textit{near isotropy of the arrival directions} of the UHECRs, and their \textit{measured flux levels}.

\subsection{Isotropy and runaway stars}

Recent measurements by the Auger observatory reveal the near-isotropy in the arrival directions to our planet of cosmic rays possessing energies above $10^{18}$ eV \cite{12, 13}. This result is incompatible with all the current conventional models about the UHECRs.

\textit{From my point of view, this nearly isotropic distribution of arrivals can be explained in terms of the tunnel ionization acceleration mechanism that I am proposing when it is combined with the workings of the massive galactic runaway OB stars that spread through the galactic halo}. In this view, these stars, after being individually expelled from the galactic disk at high speeds and in random directions, eventually undergo SN explosions, with a good fraction of them expected to leave behind electrically charged black holes, which at some stage begin accelerating cosmic rays.

In the supplementary section, the \textit{expulsion mechanism} of these stars from the galactic disk and the \textit{abundance} of their SN remnants in the galactic halo are reviewed. In that section, the \textit{long lifetimes} and \textit{near invisibility} of the expanding shells of most of them are also analyzed.

Given the randomness of the ejection direction and speeds of the runaway stars as they are expelled from the galactic disk, it is understandable that \textit{we are surrounded from all directions} by a swarm of SN-exploded massive stars and that from many of them, \textit{the Earth is receiving UHECRs with a near isotropy in the directions of their arrival}, without the need of terrestrial, solar or galactic magnetic fields to isotropize their angular distribution.

\subsection{The flux of galactic UHECRs}

The purpose in this section is to find out if the combination of this new acceleration process with the CBHs left behind by the SN explosions of the massive galactic runaway stars leads to a reasonable value for the flux of UHECRs reaching our planet.

In order to proceed in this direction, it is convenient to define in advance some \textit{average quantities} related to the galactic CBHs, as well as their symbols. Thus, $\mathcal{R}_{sn}$ is the SN explosion rate in the Galaxy ($\sim 1 / 30$ yr), $\zeta$ is the fraction of SN explosions that end up as CBHs, $\mathcal{N}_{actv}$ is the number of CBHs actively accelerating UHECRs, $\mathcal{T}_{actv}$ is the accelerating lifetime of these active CBHs, and $R_{cr}$ is the rate of acceleration of individual UHECRs by a given CBH.

If $\mathcal{R}_{sn}$ and $\mathcal{N}_{actv}$ and the flux of UHECRs in the Galaxy have remained nearly stable at least during the last few million years, then we must have that, on average, in any period of time equal to $\mathcal{T}_{actv}$, a number equal to $\mathcal{N}_{actv}$ of fresh SN has to be progressively generated in order to replace those that are becoming inactive, that is:
\begin{displaymath}
\mathcal{T}_{actv} \zeta \mathcal{R}_{sn} \approx \mathcal{N}_{actv}.
\end{displaymath}
This relation allows us to obtain the average number of years that a CBH lasts accelerating UHECRs:
\begin{equation}
\mathcal{T}_{actv} \approx \frac{\mathcal{N}_{actv}}{\zeta \mathcal{R}_{sn}}
\label{eq.10}
\end{equation}
For our galaxy, this gives
\begin{displaymath}
\mathcal{T}_{actv} \approx \frac{30 \mathcal{N}_{actv}}{\zeta} \mbox{ yr.}
\end{displaymath}
In view of the present unavailability of information about the real number of galactic BHs with diluted expanding shells, their positions, and much less about their electric charges, it is necessary to resort to a very simplistic model in order to obtain a quantitative estimate of the flux of UHECRs in a certain energy range.

\subsection{A toy model}

\textit{One thousand} charged compact stellar objects, with \textit{equal initial charges of} $10^{18}$ C are assumed to be uniformly distributed, on average, within a spherical galactic halo which has a radius of 50,000 lyr. This radius is similar to that of the halo of the Milky Way. It is assumed in this model that a new object of this type is formed every 50 years. The purpose here is to calculate the flux of UHECRs \textit{at the center of the galaxy} for the particle energies corresponding to the diminishing charges of these stellar objects, as they go down from $10^{18}$ to $10^{17}$ C. Although the flux of these particles should be less at the radial distance equivalent to that of the Earth from the galactic center ($\sim 27,200$ lyr), these two fluxes should differ by only a moderate factor.

Therefore, in this case, we have $\mathcal{N}_{actv} = 1,000$, $\zeta \mathcal{R}_{sn} = 1 / 50$ yr, and consequently, the following accelerating lifetime is obtained:
\begin{displaymath}
\mathcal{T}_{actv} \approx 50,000 \mbox{ yr}.
\end{displaymath}
In terms of elementary charges, the initial charge is $Q_{bh} = 10^{18}$ C $\approx 6.241 \times 10^{36} e$, and this charge will be reduced to $10^{17}$ C after $0.9 \times 6.241 \times 10^{36}$ accelerating events. If for simplicity, it is assumed that this 90 \% charge reduction takes place during 90 \% of $\mathcal{T}_{actv}$, then it is obtained that the rate of accelerated CR is $R_{cr} \approx 1.25 \times 10^{32} \mbox{ yr}^{-1}$ (or $\sim 4 \times 10^{24} \mbox{ s}^{-1}$). As shown in table \ref{tab.1}, the energies of the UHECRs go from $\sim 7.8$ EeV when $Q_{bh} = 10^{18}$ C to $\sim 2.5$ EeV, when it is $10^{17}$ C.

\begin{figure}[!h]
  \centering
  \includegraphics[height=6in]{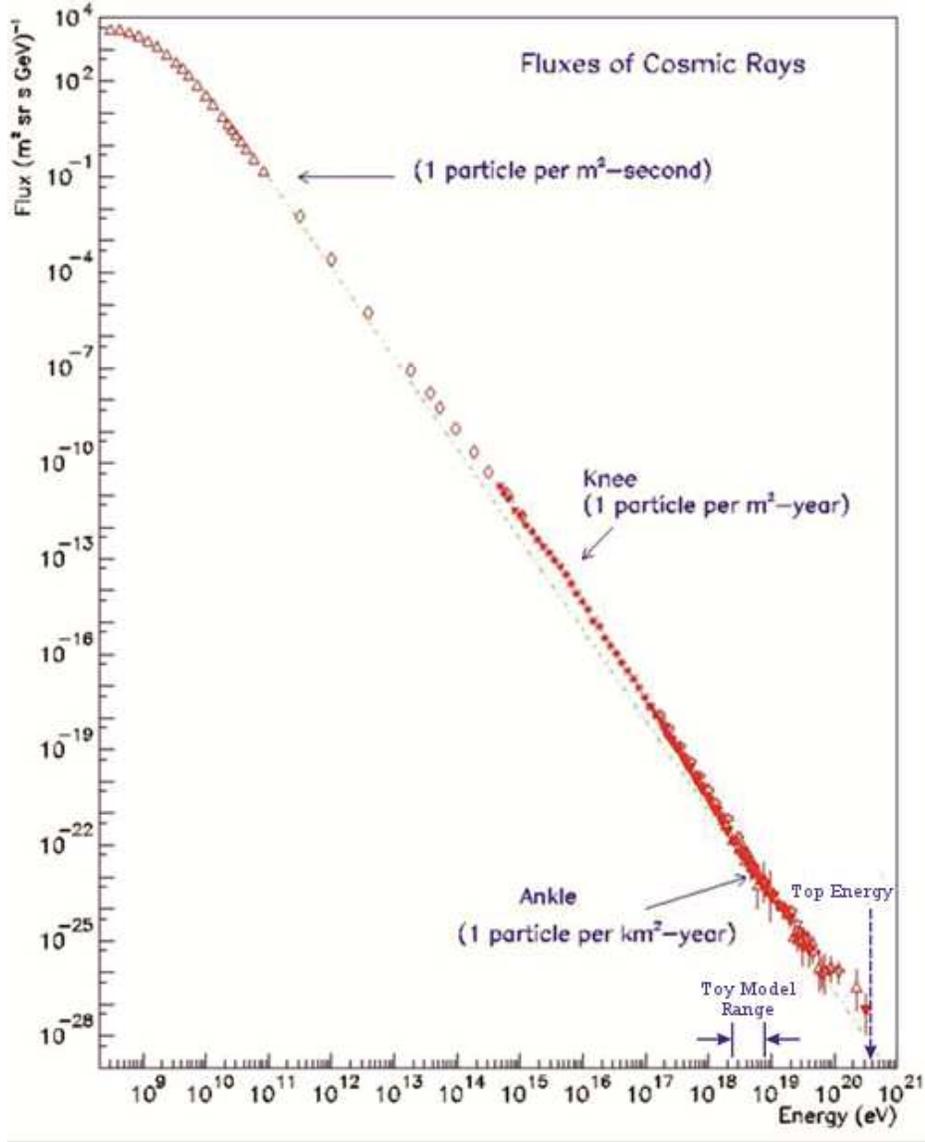}
  \caption{In this energy spectrum of the CR, the predicted upper energy limit $W_{eV}^{Max}$ is indicated by the dashed vertical arrow. The energy range covered by the toy model is indicated in the lower part of the figure.}
\label{fig.3}
\end{figure}

Since the volume of this galactic halo is
\begin{displaymath}
V_{halo} = \frac{4}{3} \pi r_{halo}^{3} \approx 5.24 \times 10^{14} \mbox{ lyr}^{3},
\end{displaymath}
then the average volume per actively accelerating CBH is $\Delta V_{cbh} \approx 5.24 \times 10^{11} \mbox{ lyr}^{3}$.  If in addition, it is assumed that these UHECRs are minimaly deviated by the galactic magnetic fields, with their paths remaining nearly straight lines, then the flux contributed at the center of the galaxy by a single active CBH situated at the radial distance $r$ from this center is given by
\begin{displaymath}
\Delta \phi = \frac{R_{cr}}{4 \pi r^{2}}.
\end{displaymath}
So, in order to find the total flux at the center, it is necessary to add the flux ontributions from all the active CBHs within the halo. Since this type of calculation can be somewhat cumbersome, even if these CBH are allocated in a cubic lattice, the required discrete summation can be better approximated by an integration over a uniform and continuous distribution of sources in the halo. To this end, we first write down the flux contribution $d \phi$ of the element of volume $dV$, situated at the distance $r$ from the center as:
\begin{equation}
d \phi = \Delta \phi \times \frac{dV}{\Delta V_{cbh}} = \frac{R_{cr}}{4 \pi r^{2}} \times \frac{dV}{\Delta V_{cbh}}
\label{eq.11}
\end{equation}
By taking $dV$ as a spherical shell of radius $r$ and thickness $dr$; that is, $dV = 4 \pi r^{2} dr$, then the following simple form is obtained:
\begin{displaymath}
d \phi = \frac{R_{cr}}{\Delta V_{cbh}} dr.
\end{displaymath}
Therefore, the total flux of UHECRs with energies between $2.5 \times 10^{18}$ and $7.8 \times 10^{18}$ eV is given by
\begin{equation}
\phi_{gc} = \int_{0}^{r_{halo}} \frac{R_{cr}}{\Delta V_{cbh}} dr = \frac{R_{cr}}{\Delta V_{cbh}} r_{halo}
\label{eq.12}
\end{equation}
When the previously defined values of $R_{cr}$, $\Delta V_{cbh}$, and $r_{halo}$ are substituted in this equation, and the value 1 lyr = $9.461 \times 10^{12}$ km is used, the following result is obtained for the flux of these energetic particles at the galactic center:
\begin{equation}
\phi_{gc} = 0.133 \mbox{ /km}^{2} \mbox{-yr}
\label{eq.13}
\end{equation}
This flux value is surprisingly similar to the experimental value measured at the Earth for this energy range, as can be seen in the Ankle region of the cosmic ray spectrum shown in figure \ref{fig.3} (\texttt{http://apcauger.in2p3.fr/Public/Presentation}).

Undoubtedly, this quantitative closeness to the measured flux of these particles at Earth constitutes a compelling evidence in favor of the tunnel ionization acceleration mechanism proposed in this work and not just an additional compatibility. This is so considered, in spite of the improvements that would have to be made to the halo model and to the calculations if complete information about the sources and their locations were available.

\subsubsection{The flux of UHECRs outside the halo}

Still within the toy model, but now without the need of the approximation of the continuous distribution of sources, the following expression is obtained for the flux of UHECRs at places located well outside the galactic halo:
\begin{equation}
\phi_{out} (r) \approx \frac{\mathcal{N}_{actv} R_{cr}}{4 \pi r^{2}}, \hspace{5cm} \mbox{for } r \gg r_{halo}
\label{eq.14}
\end{equation}
This is so because all the sources are nearly at the same distance from the remote observational point. For example, for $r = 10^{6}$ lyr ($r = 20 r_{halo}$) and for the same previous sources, we obtain:
\begin{displaymath}
\phi_{out} (\mbox{Mlyr}) \approx 1.11 \times 10^{-4} \mbox{/km}^{2}\mbox{yr } \approx 8.4 \times 10^{-4} \phi_{gc}
\end{displaymath}
Conversely, we can think of a similar toy-model halo for Andromeda, but this time with a smaller value of $\mathcal{N}_{actv}$ due to its smaller rate of SN explosions when compared to our galaxy. At a distance of $\sim 2.5 \times 10^{6}$ lyr from us, the flux reaching the Earth from that object would be $\phi_{out} (\mbox{2.5 Mlyr}) \lesssim 1.36 \times 10^{-4} \phi_{gc}$, which is less than 3 particles/$\mbox{km}^{2}$-millenium. Even smaller fluxes are received from other nearby galaxies possessing star-forming regions, but which are located at greater distances to us than Andromeda.

These results clearly indicate that the bulk of the UHECRs reaching our planet are generated by stellar objects located in our galaxy, and that their flux decay to very low levels at distances of only a few Mlyr from it. This does not prevent that a tiny fraction of all these UHECRs has an extragalactic origin.

\subsubsection{The Ankle}

This feature is a small irregularity in the energy spectrum of figure \ref{fig.3}, located at $\sim 4 \times 10^{18}$ eV, where the intensity power-law suddenly changes from $\propto E^{-3.3}$ to $\propto E^{-2.7}$, a phenomenon described as the hardening of the energy spectrum \cite{14}.

For many years, the interpretation of the Ankle has been that it marks the transition of the dominance between the galactic and extragalactic CRs. The idea is that once the gyration radius $r_{g}$ of the charged particle approaches the accelerator size, it becomes increasingly difficult to magnetically confine the CR in the galactic acceleration region, and thus to continue the acceleration process up to higher energies (beyond the Ankle) through Fermi's first-order acceleration mechanism \cite[p. 1677]{15}. Therefore, the necessary conclusion about the extragalactic origin of the particles with energies beyond this feature. For the estimated galactic fields of a few tens of $\mu$G $r_{g}$, these dimensions are reached when the proton energy is in the Ankle region \cite{16}. This transitional interpretation of the Ankle has been further strengthed because of the incompatibility of the near isotropy of arrivals of these particles with an origin for them from sources in the galactic disk \cite{12, 13}.

\textit{From my point of view, the Ankle is again originated by the sharp cutoff of the acceleration of CRs by the shock fronts of the SN blast-waves in the halo, but this time, it operates in combination with the continuation of the new acceleration process of CRs from energies below the Ankle to energies above it, all the way through the highest measured energies}. However, it is not assumed that these two processes can take place simultaneously in the same individual SNR but that, in general, they take place in individuals having different levels of evolution. Thus, we have that the new acceleration model being proposed here \textit{is also compatible with the Ankle feature}, but again, with the great majority of the UHECRs coming to Earth from our own galaxy, and not from extragalactic sources.

In the supplement, some considerations are made about the non-incompatibility of this new acceleration scheme with the rest of the energy spectrum of the CRs above the Ankle, all the way through the top energy of the UHECRs and also about the possibility of the sources of the particles in the ``toe" region of the spectrum \textit{of being closer to Earth}, on average, than the sources of the other UHECRs.

\subsection{The fate of the ionized electrons}

The dynamical description of the dissociated electron as it spirals towards the CBH requires a full relativistic treatment, because of the speeds and the colossal electric and gravitational fields involved. Presently, this type of theory lies well outside my reach.

However, it is possible to make an order of magnitude calculation of the energy $W_{e}$ available to the electron as it falls from the ionization radial distance $r_{sh}$ to the Schwarzchild or horizon radius $r_{schw}$ of the BH. In eV, the non-relativistic value of this energy is given by:
\begin{equation}
W_{e} \approx 6.2415 \times 10^{18} \left( \frac{Q_{bh} e}{4 \pi \epsilon_{0}} \right) \left[ \frac{1}{r_{schw}} - \frac{1}{r_{sh}} \right] \mbox{ eV/J}
\label{eq.15}
\end{equation}
For the same CBH chosen above, with $Q_{bh} = 10^{18}$ C and $M_{bh} = 10 \mbox{ M}_{\odot}$, from table \ref{tab.1}, we obtain $r_{sh} \approx 1.164 \times 10^{6}$ km On the other hand, for a BH of this mass, the Schwarzschild or horizon radius given by $r_{schw} \approx 2.95 M_{bh} / \mbox{M}_{\odot}$ km, we get $r_{schw} \approx 29.5$ km. When these values are substituted in equation \ref{eq.15}, we obtain $\mathbf{W_{e} \approx 3 \times 10^{23}}$ \textbf{eV}. When this value is compared with the proton repulsive energy of $7.78 \times 10^{18}$ eV for this CBH case, we find that \textit{the energy available to the electron is near $3.9 \times 10^{4}$ times higher than the energy of the accelerated proton itself!}

If some magnetic fields were present around the CBH, then there is the possibility that a fraction of this energy available to a single electron will be transformed into prodigious amounts of photons due to the ultra-relativistic synchrotron radiation of the accelerated charge, with the energy of these photons distributed into a continuum energy spectrum spanning from less than one eV up to the GeV and TeV ranges. Thus, for example, within only 1 \% of $W_{e}$, there is sufficient energy to produce a combination of more than $10^{20}$ photons in the eV band, $10^{17}$ in the keV, $10^{14}$ in the MeV, $10^{11}$ in the GeV, and $10^{8}$ in the TeV bands, emitted in different directions. Thus, there is also the prospect for the tunnel ionization acceleration mechanism of explaining the origin of the point sources observed in GeV and TeV gamma-ray astronomy \cite{17, 18}, as well.

\subsection{A second toy model}

Let us again consider the same CBH, but this time at a distance of $10^{4}$ lyr from Earth and generating photons according to a fictitious energy spectrum, taken from the above simplistic combination, in each one of its UHECR acceleration events. Assuming as before, that the rate of the accelerated cosmic rays is $R_{cr} \approx 4 \times 10^{24} \mbox{ s}^{-1}$, then we obtain that the \textit{flux of cosmic rays} produced by that source at the Earth would be about $3.6 \times 10^{-17} \mbox{/m}^{2} \mbox{s}$, while those of the \textit{photons} would be $3.6 \times 10^{3}  \mbox{/m}^{2} \mbox{s}$ for the eV, $3.6  \mbox{/m}^{2} \mbox{s}$ for the keV, $3.6 \times 10^{-3}  \mbox{/m}^{2} \mbox{s}$ for the MeV, $3.6 \times 10^{-6}  \mbox{/m}^{2} \mbox{s}$ (that is, $1.1 \times 10^{2} \mbox{/m}^{2} \mbox{yr}$) for the GeV, and $3.6 \times 10^{-9}  \mbox{/m}^{2} \mbox{s}$ for the TeV bands.

The flux obtained for the GeV photons with this crude and naive model is not too discordant with the fluxes required by the Fermi-LAT orbiting telescope for imaging some distant galactic point sources emitting in this band \cite{17}. Something similar can be said about the flux obtained for the TeV photons in relation to fluxes required by the ground-based telescopes, which record Cherenkov radiation from TeV gamma rays in the upper atmosphere, to image some galactic point sources \cite{18}. Thus, I think that is is safe to assert that this new acceleration model of UHECRs is \textit{not incompatible} with the experimental observation related to galactic point sources in GeV and TeV Astronomy. However, it will be necessary to wait for the calculations of realistic energy spectra of the photons generated by ultra-relativistic synchrotron radiation phenomena of the accelerated electrons in order to see if they lead to reasonable agreements with observations. This is something which I presently cannot do.

On the other hand, the photon flux in the eV band obtained through this toy model is rather high. Even if it were \textit{several orders of magnitude lower}, it would be possible to obtain the image of a feeble \textit{non-thermal} point source if a large area \textit{optical} telescope were aimed toward one of the known galactic GeV or TeV gamma-ray point sources during an extended time. This is in the case that the acceleration mechanism of UHECRs proposed here were actually operating in nature.

\section{Conclusions}
\label{sec.6}

By taking as real the filling of the niche predicted by the K-N theory for the acquisition of colossal amounts of electric charge by recently formed stellar BHs and by assuming that the rapid loss of this charge is prevented by the expanding shock front of the stellar explosion, and by combining these with the tunnel-ionization acceleration mechanism, a sizable list of compatibilities with the experimental observations is obtained, with a couple of them in near quantitative agreement. In addition, from this new acceleration scheme, a picture clearly emerges in which the bulk of the CR are being accelerated in our own galaxy \textit{in their entire energy spectrum} (without excluding the possibility that a minority of them come from sources outside it). This is in sharp contrast with the prevalent view that all the UHECRs with energies above the Ankle feature have to come from extragalactic sources.

\end{document}


\maketitle

\section{Two types of force acting on neutral atoms}
\label{sec:1}

The attractive force of the CBH on a neutral atom consists of two main components. The first of them is the well-known gravitational force, which is proportional to $r^{-2}$. The second force is of electric nature: \textit{the strength of the radial electric field of the CBH polarizes the atom and its radial gradient attracts the induced dipole.} The combined effect gives a force proportional to $r^{-5}$. Complete simulations were carried out with and without the inclusion of this last force, but not noticeable differences between the two cases were observed. This occurs because the ionization process takes place at radial distances much larger than those where the dipolar force becomes comparable to, or greater than, the gravitational force. For this reason, only the gravitational force was considered in this work.

\section{The occurrence of the BH charge}
\label{sec:2}

Some preliminary arguments for the long-term durability of the electric charge of BHs were advanced in the main article. Here, I want to comment about some potential sources for these BH charges, keeping in mind that their existence was one of the \textit{working hypotheses} which have permitted to reach an important list of compatibilities with observational measurements that give credibility to the new acceleration model. However, this credibility should not be affected by the degree of plausibility of some of the following considerations.

As I see it, for the creation of the \textit{Kerr-Newman} charge there are two possible routes: either it is created \textit{internally}, as part of the process of the star explosion or it is created \textit{externally}, after the explosion and external to the residual BH.

\subsection{Internal mechanism}
\label{sec:2.1}

After more than fifty years of theory, advanced computer calculations and observations, the process by which a star terminates its long life so violently within seconds is not yet understood. The majority of the core collapse simulations do not lead to star explosions, and those that do at best obtain a small fraction of the characteristic energies observed. Even less is understood about other aspects of SN explosions, such as nucleosynthetic yields as a function of the stellar progenitor, the branch map connecting the progenitor to either a neutron star or a stellar-mass BH final state, the explosion morphology and the spatial distribution of the ejected elements, the temporal light curves and their spectra, and a host of other facts surrounding the SN phenomenon \cite[p. 245]{Burr13}. \textit{Thus, presently, there are no theoretical or observational elements that allow us to sustain that the central remnant of a SN explosion can or cannot acquire an important electrical charge through this internal route.}

\subsection{External mechanism}
\label{sec:2.2}

These mechanisms have to do with material which, after being expelled from the star during the SN explosion, returns to it in the form of small rocks and dust particles after having been exposed to a bath of energetic photons and to intense variable magnetic fields. The formation of dust particles in the SN remnants (SNR) is widely commented in the literature; for example, in the reference \cite[p.L33]{HLR98}, it is mentioned that grains of graphite and refractory oxides are condensed in SN ejecta. For these particles, the gravitational pull of a CBH could be much stronger than the electric repulsion. For instance, in our customary stellar CBH, I obtain that the \textit{weight} of a particle which has lost one electron is 10 times larger than the \textit{electric repulsive force} when the mass of the particle is just $\sim 10^{-8}$ gr. Each falling particle like this one would contribute to the increase of charge of the BH by $+e$.

There are two different observational cases of this type of processes: The \textit{fallback material} and the \textit{reverse shock wave}. The first of them occurs when there is a large amount of the ejected material during a SN explosion which does not reach the escape velocity and falls back to the proto-neutron star. This occurs for sufficiently large progenitor stars ($\sim 20$ to 40 $\mbox{M}_{\odot}$). This fallback material then turns the young neutron star into a stellar BH. See for example \cite[p.411]{MWH01}. In relation to the second case, not long after the SN explosion, the deceleration of the outer blast wave causes the formation of a reverse shock as the inner ejecta are forced to decelerate. Thus, this phase of the explosion is characterized by the simultaneous presence of both forward shock (blast wave) heating and sweeping up the interstellar medium and the reverse shock  heating ejecta. Eventually, after \textit{several thousand years}, the reverse shock moves into the central region of the SNR, and after what may be extensive reverberations, it disappears. By this times, several times the ejected mass has been swept up by the blast wave \cite{RGB11}.

There exists another natural phenomenon that could help the BH to build up a large electric charge. It has to do with the observational fact that the majority of the compact stellar objects left behind by the SN explosion receive a \textit{kick-velocity} of several hundreds of km/s \cite{HBZ01, SRTT10}. This is perhaps caused by the asymmetry of the explosions. Thus, we have the potential scenario of a neutron star or a BH moving at high speed through clouds of material that was previously ejected by the same explosive event. This could lead to an enhancement of the electric charging of the BH.

\section{Electrically charged neutron stars}
\label{sec:3}

For less massive progenitor stars whose SN explosion give rise to \textit{neutron star}, instead of BH, it is logical to think that there must always be some amount of fallback material. This idea is consistent with the SN 1987A case. It is generally agreed that this stellar explosion initial produced a neutron star of mass $\sim 1.4 \mbox{M}_{\odot}$ and that there might have been $\sim 0.1 \mbox{M}_{\odot}$ of fallback material onto this neutron star, although it is not yet known if it remained as a neutron star or if a BH was formed \cite[p.425]{MWW01}. Consequently, the possibility of acquiring large electric charges by fallback should not be restricted to stellar BH, but should also be shared by neutron stars; although in this case, I am not aware of something similar to the Kerr-Newman relation of the CBHs relating the electric charge of the neutron star to other of its parameters. For this reason, in this work, the term CBH refers both to \textit{charged BH} and to \textit{charged neutron stars}.

\section{The expulsion mechanisms of the runaway stars}
\label{sec:4}

About 20 \% of all massive stars of the Milky Way have unusually high velocities and it is believed that the majority of them were born in dense and relatively low-mass young star clusters situated in the galactic plane \cite{FPZ11}. Their high speeds of ejection (in extreme cases exceeding 700 km/s) and their random directions have been explained on the basis of \textit{two mechanisms}: the first of them consists of a \textit{three-body gravitational slingshot effect} that typically results in the less massive star being ejected from the cluster. In the other mechanism, the \textit{SN explosion of the larger member} of a binary-star system leaves the other star loose and with a velocity exceeding the escape speed of the cluster \cite{HBZ01, LO07, RHM01, SR08}.

By using the well-known formula
\begin{displaymath}
T \approx \left( \frac{\mbox{M}_{\odot}}{M} \right)^{2.5} \times 10^{10} \mbox{ yr}
\end{displaymath}
for the \textit{lifetime of a main sequence star} of mass $M$, it is found that for stars with $M = 8$, 20, and $40 \mbox{ M}_{\odot}$, for example, their corresponding lifetimes are $T \approx 55.2$, 4.6, and $1.0 \times 10^{6}$ yr, respectively. 

By assuming an intermediate ejection speed of 300 km/s = $10^{-3} \mbox{ }c$, it is also found that, in addition to their orbital displacement around the galaxy, these stars move through the galactic potential along trajectories of lengths $L \approx 55.2$, 5.6, and $1.0 \times 10^{3}$ lyr, respectively, \textit{before they end their lives as SN} (a stellar mass of 8 $\mbox{M}_{\odot}$ is about the minimum required for a SN explosion) \textit{and some of these stars should be capable of reaching the outer fringes of the galactic halo.} On the other hand, the heavier runaway stars are expected to explode closer to the galactic plane. 

In addition, there is observational evidence that there may be as many as several hundreds of young B-type stars (capable of undergoing a SN explosion) that were born directly in the galactic halo, or born \textit{in situ} \cite{RHK99}.

\section{The population of accelerating CBHs in the galactic halo}
\label{sec:5}

According to my preliminary argument advanced in the main article, the acceleration of UHECRs takes place after the expanding shell of the SNR has been sufficiently weakened. Since the number of CBHs \textit{actively accelerating} UHECRs must be smaller than the number of SNR shells, the number of these shells can be used as an upper bound for the number of the more difficult to observe active CBHs. Now, these expanding shells are known to be very long-lived, reaching up to $10^{6}$ yr \cite{CBh98}. Since in our galaxy, a SN outburst occurs every 30-50 yr on average (by neglecting those Crab-like SNRs which do not contain expanding shell structures), the total number of shell remnants should be $2 - 3 \times 10^{4}$. 

Since from radio and X-ray studies, only about 230 of these objects are known to exist, it is then clear that most of the shell SNRs are ``missing" \cite{KKS06}. The in-depth study of shell SNRs has depended largely on observations in the radio regime. However, radio telescope surveys easily overlook those shells of low surface brightness ($\Sigma$) or small angular size \cite{CBh98}. One illustrative case of this is a rapidly expanding old shell ($\sim 80$ km/s) believed to be originated by a SN explosion that occurred in the outermost fringes of our galaxy some $\sim 3 \times 10^{5}$ yr ago. This SNR shell was revealed by a high resolution $\mbox{H}_{\mathrm{I}}$ (21 cm) observation, but it is essentially invisible in all other radio bands \cite{KKS06}. This problem of the missing shells is originated, to a great extent, by the dependence of the surface brightness of the shell on the normal distance $\abs{z}$ of the SN explosion to the galactic plane.

An observable radio-astronomical property of SNR shells is that, on the average, they are brighter on the side closer to the galactic plane than on the more distant side. The quantitative analysis of a number of SNR shells has permitted to establish the following empirical formula for the surface brightness (at 408 MHz):
\begin{displaymath}
\Sigma = 1.25 \times 10^{-15} t^{-6/5} e^{- \abs{z} / 359},
\end{displaymath}
with the age $t$ measured in yr and the height $z$ in lyr. At the same time, this $z$-dependence of $\Sigma$ is partially caused by the $z$-dependence of the density of the \textit{diffuse galactic medium} given by the also empirical formula
\begin{displaymath}
\rho (z) \approx \rho (0) e^{- \abs{z} / 587},
\end{displaymath}
with $z$ again measured in lyr \cite{CL79}. In evaluating the factor $e^{- \abs{z} / 359}$ for $\abs{z} =$ (1, 2.5, 5, 10, and 20) $\times 10^{3}$ lyr, the values of this factor are 0.06, $9.5 \times 10^{-4}$, $8.9 \times 10^{-13}$ and $6.4 \times 10^{-15}$, respectively. Thus, the amount of galactic gas and dust particles to be swept and heated up by the expanding shell rapidly reduces to negligible values with increasing $\abs{z}$, and so does $\Sigma$. \textit{This explains the practical invisibility of the SNR shells at large galactic heights since their extremely weak signals are buried by the cosmic electromagnetic noise.} However, observations with more advanced radio telescopes and with space-based X-ray telescopes have permitted the discovery of a number of very faint or small angular-size SNR-shells, \textit{clearly indicating that there are many more of these objects awaiting for their discovery} \cite{CBh98, KKS06}. As the number of known SN expanding shells increases, so will the potential number of CBHs increase.\footnote{My introduction of SNRs in the galactic halo as sources of UHECRs, and having massive runaway stars as progenitors, appears to be a novel idea. All that I have met in my readings of the related literature is the deeply-rooted idea that if these CR cannot come from the galactic disk, then they have to come from extragalactic objects.}

\section{Above the Ankle}
\label{sec:6}

The scheme consisting of the combination of the tunnel-ionization acceleration mechanism and the distribution of CBHs in the galactic halo is also \textit{not incompatible} with the rest of the energy spectrum of CRs above the Ankle. For example, within this scheme, the measured events in the ``toe" region of the energy spectrum ($E \gtrsim 5 \times 10^{19}$ eV) require the largest CBHs in the halo, with their full charge.

For the flux of CRs coming from these CBHs, there are \textit{two opposite tendencies}, one for \textit{increasing} this flux and the other for \textit{decreasing} it. Their descriptions are as follows:
\begin{enumerate}
\item The value of the cosmic-ray acceleration rate parameter $R_{cr}$ obtained in the main article was an average value, but since the ionization cross section of a CBH is proportional to its charge\footnote{As is explained in the main article and graphically shown in figure 1 (A), the ionization of the H atoms takes place on a spherical shell of radius $r_{sh} \approx 8.8 r_{B}$, with $r_{B} = a_{0} (Q_{bh} / q_{p})^{1/2}$, then the cross section of the CBH can be defined as $A_{sh} = \pi r_{sh}^{2}$, hence $A_{sh} \approx \pi (8.8 a_{0})^{2} Q_{bh} / q_{p}$. If the BH charge is expressed as $Q_{bh} = 10^{n}$ C $\approx 6.242 \times 10^{n + 18} q_{p}$ and with $a_{0} \approx 5.29 \times 10^{-11}$ m, then the cross section becomes: $A_{sh} \approx 4.25 \times 10^{n} \mbox{ m}^{2}$.}, these stellar objects \textit{will tend to have the largest values} of $R_{cr}$, and we also need to take into consideration that these large objects will remain in regions near the galactic plane, with \textit{higher densities of galactic gas particles} to be ionized and accelerated.

\item These very large CBHs come from the most massive runaway progenitors which, according to the Salpeter Initial Mass Function ($N (M) \propto M^{-2.35}$), \textit{are the fewest}. Besides that, \textit{they undergo SN explosions before reaching high values of} $\abs{z}$ from the plane and the large majority of their emitted CRs are rapidly stopped by the same dense interstellar medium that surrounds them. 
\end{enumerate}

Since it is unlikely that these two tendencies will exactly balance each other for all the directions, \textit{no isotropy of arrivals has to be expected} for these top-energy CRs in the toe region. Instead, some degree of clustering (above or below the galactic plane) is to be expected in those directions of the galactic plane pointing toward recent, large-star formation regions located in the vicinity of the Sun.

So, in sharp contrast with the general view that the most energetic UHECRs have to come from remote extragalactic regions, this new  acceleration scheme leads to the view that the average distance of their sources to us is shorter, at a galactic scale, than those of the sources of the other above-the ankle energy ranges. This is because the more distant giant sources are ``out of sight" from Earth, unless these sources have been created \textit{in situ}, in the galactic halo. Perhaps a similar type of considerations could be applied to the flux of the UHECRs with energies in the region between the \textit{ankle} and the \textit{toe} of the spectrum.

\section{The acceleration and abundances of ``galactic'' CRs and the heating and composition of the solar corona}
\label{sec:7}

It was mentioned in section \ref{sec:2.1}, that the abundances of the galactic CRs, which possess energies in the GeV-TeV range and are assumed to be accelerated by the shock front of the SNRs, seem to be controlled by atomic paramters such as the \textit{first ionization potential} (FIP), with the lower FIP chemical elements being systematically favored relative to the higher FIP ones. This phenomenon cannot be explained in terms of the DSA acceleration mechanism. However, it is compatible with the acceleration mechanism that is being introduced in this work.

There also exists an extremely similar FIP bias in the compositions of the solar corona, in the solar wind and in solar energetic particles \cite{MDE97}. At the same time, the cause of the much higher temperature of the solar corona (over $10^{6}$ K) than that of the solar chromosphere (under 6000 K) is still an unsolved mystery in solar physics, after more than seven decades of the discovery of this phenomenon.

Since the \textit{tunnel ionization acceleration mechanism} also depends on atomic, and not on nuclear properties, which are closely related to the FIPs, there is the possibility that the tunnel ionization mechanism of \textit{neutral and nearly stationary} atoms is also playing important roles at the shock fronts of SNRs for the \textit{acceleration of the galactic CRs} (thus complementing the DSA acceleration mechanism, or perhaps challenging it!), and at the solar chromosphere for the \textit{heating and composition-building} of the corona. The two cases potentially caused by strong transient electric fields and \textit{in single-acceleration events}, with these fields being generated by fast traveling and time-changing magnetic fields in extensive regions around the atoms, as in the case of evolving solar magnetic loops. In these tenuous and hot environments, the electric fields could potentially tunnel-ionize and accelerate \textbf{\textit{highly excited neutral atoms}} more easily than those atoms in their ground states. For the solar case, this mechanism could be capable of injecting energetic ions into the corona, with element-dependent efficiencies.

These are two additional compatibilities of the tunnel-ionization acceleration mechanism with astrophysical observational facts. The \textit{ultimate case} would be if the tunnel ionization acceleration mechanism were also the \textit{ingredient that is missing} in the different types of computer simulations of SN explosions, which until now have not been fully successful!\footnote{The big difference between the outwardly acceleration of already formed ions and the tunnel-ionization acceleration mechanism is that in the first case, these charged particles are accelerated when the electric field is still weak, and usually cannot reach very high energies, while in the second case, the neutral atoms can accede to the extremely high field regions, and upon their tunnel-ionization, their positive and negative parts can be accelerated by these fields along extended paths, acquiring enormous kinetic energies at the end. This is apart from the element-dependence difference between the two processes.}